\documentclass[reprint,twocolumn,notitlepage,prl,superscriptaddress,showpacs]{revtex4-2}
\usepackage{hyperref}
\usepackage[usenames,dvipsnames]{color}
\usepackage{amsmath}
\usepackage[english]{babel}
\usepackage{amssymb}
\usepackage{graphicx}
\usepackage{epstopdf}
\usepackage{comment}
\usepackage{soul}
\usepackage{marvosym}
\usepackage{mathrsfs}
\usepackage{amsfonts}
\usepackage{tabularx}
\usepackage{multirow}

\let\oldbibitem\bibitem
\renewcommand{\bibitem}{%
  \renewcommand{\doi}[1]{doi: ##1}
  \let\bibitem\oldbibitem
  \oldbibitem
}

\begin{document}

\title{Long-wavelength phonon dynamics in incommensurate \texorpdfstring{Bi$_2$Sr$_2$CaCu$_2$O$_{8+\delta}$}\. crystals by Brillouin light scattering spectroscopy}
\author{B. D. E. McNiven}
\author{J. P. F. LeBlanc}
\author{G. T. Andrews}
\email{tandrews@mun.ca}
\affiliation{Department of Physics and Physical Oceanography, Memorial University of Newfoundland and Labrador, St. John's, Newfoundland \& Labrador, Canada A1B 3X7}

\date{\today}
\begin{abstract}
Room temperature phonon dynamics in crystals of the high-$T_c$ superconductor Bi$_2$Sr$_2$CaCu$_2$O$_{8+\delta}$ were probed using Brillouin light scattering spectroscopy. Eight distinct acoustic modes were observed and identified, including two quasi-longitudinal bulk modes and four quasi-transverse bulk modes. A peak at a frequency shift of $\sim95$ GHz with behaviour reminiscent of an optic phonon was also observed in the spectra.  The existence and nature of these modes is a manifestation of the incommensurate structure of Bi$_2$Sr$_2$CaCu$_2$O$_{8+\delta}$ and suggests that it may be categorized as a so-called composite incommensurate crystal comprised of two weakly interacting sublattices.  A mass ratio of $m_1/m_2=0.36$ obtained from the two measured quasi-longitudinal acoustic velocities led to sublattice assignments of Bi$_2$Sr$_2$O$_4$ and CaCu$_2$O$_4$.  The Brillouin data also places an upper limit of $\sim10$ GHz on the crossover frequency between commensurate and incommensurate phonon dynamics. 
\end{abstract}

\maketitle

\section{Introduction}
Quasiparticle dynamics in incommensurate systems are not well understood, with the scarcity of knowledge being particularly acute for long-wavelength acoustic phonons. While the ``usual" phonon dispersion for a typical commensurate crystal consists of 3$N$-3 optic and 3 acoustic phonon modes, where $N$ is the number of atoms in the unit cell, theoretical studies predict modifications to the phonon dispersions of incommensurate crystals due to their unique structure \cite{Axe1982,Currat2002,Dzugutov1995,Finger1982,Janssen2002,Radulescu2002}. Moreover, complementary experimental work on acoustic phonon dynamics in such systems is limited despite the fact that many have incommensurate phases at ambient temperatures, making them prime candidates for laboratory investigation.  The crystalline form of high-temperature superconductor Bi$_{2}$Sr$_{2}$CaCu$_{2}$O$_{8+\delta}$ (Bi-2212) is one such material \cite{Miles1998,Kan1992,Ariosa2001}.

Experiments aimed at probing acoustic phonon dynamics in Bi-2212 reveal complex dispersion with branch crossings and the unexpected absence or presence of particular modes arising from the incommensurate structure. For example, inelastic neutron scattering results suggest that coupling of longitudinal acoustic phonons to electrons is limited to zone center phonons by branch anticrossings, and that low-energy spectral weight near a charge density wave ordering wave vector arises from an extra low-lying acoustic phonon branch emanating from a nearby superlattice modulation reflection \cite{Merritt2019}.  Possibly consistent with this second suggestion is the observation of two (instead of the usual one) longitudinal acoustic-like modes along the incommensurate direction in another neutron scattering study \cite{Etrillard2001}. Interestingly, only one longitudinal mode was measured in this direction in ultrasonic pulse-echo experiments conducted at 7.5 MHz \cite{Wang1}. These last two observations suggest an undiscovered crossover from two propagating longitudinal modes to one such mode in the 10 MHz $-$ 150 GHz frequency regime inaccessible to these two techniques. Furthermore, neutron scattering experiments conducted along the incommensurate direction in Bi-2212 detected only one transverse mode instead of the two that were expected \cite{Etrillard2001}. Additionally, Ref. \cite{Merritt2019} indicates that a minimum of eight phonon branches can be expected for Bi-2212. Given that the GHz-frequency regime remains largely unexplored, it is likely that one or more of these branches could exist in this frequency range.

This paper reports on Brillouin light scattering experiments on high-quality single crystals of Bi-2212, the need for which has been repeatedly articulated in the literature \cite{Finger1982,Etrillard2001,Etrillard2004}. By probing with sub-GHz resolution a frequency range inaccessible to ultrasonics and neutron scattering techniques, the results of this study provide new insight into the dynamics of long-wavelength hypersonic phonons in Bi-2212. In particular, the incommensurate structure of Bi-2212 is manifested in the Brillouin spectrum through the presence of multiple acoustic phonon peaks in excess of those expected for a typical crystal.  This observation places an upper limit of $\sim10$ GHz on the ``separate sublattice-to-standard single crystal" crossover frequency and suggests that Bi-2212 is more accurately described as a composite incommensurate crystal rather than a modulated incommensurate crystal. Moreover, quasi-longitudinal phonon velocities determined in this work permit identification of the two sublattices that comprise the incommensurate Bi-2212 crystal.  The damping-out, with increasing proximity to the incommensurate direction, of a previously unreported low-lying optic-like mode at $\sim95$ GHz may be another manifestation of the incommensurate character of Bi-2212.  The above observations make this the first Brillouin scattering study to reveal the incommensurate nature of Bi-2212.  

\section{Experimental Details}
Brillouin scattering experiments were performed at room-temperature using a 180$^\circ$ backscattering geometry on (001)-oriented flakes of Bi-2212 exfoliated from three parent crystals with critical temperatures of 78 K, 90 K, and 91 K.  Incident light was provided by a single-mode frequency-doubled Nd:YVO$_4$ laser emitting at a wavelength of $\lambda_{i}=532$ nm. 
The beam was horizontally (``p") polarized by use of a half-wave plate and then passed through neutral density filters to reduce the power at the sample to $\sim$10 mW to minimize heating due to optical absorption. The scattered light was collected and collimated by a high-quality camera lens with an aperture setting of $f/2.8$ and subsequently focused onto the entrance pinhole of a six-pass tandem Fabry-Perot interferometer using a $f=40$ cm lens. The pinhole diameter was set to 300 $\mu$m or 450 $\mu$m, with the former being used when the central elastic peak width needed to be reduced to reveal Brillouin peaks at very small frequency shifts. 

\section{Results and Discussion}

\subsection{Brillouin Spectra - General Features}
\begin{figure}[!t]
  \includegraphics[scale=0.79]{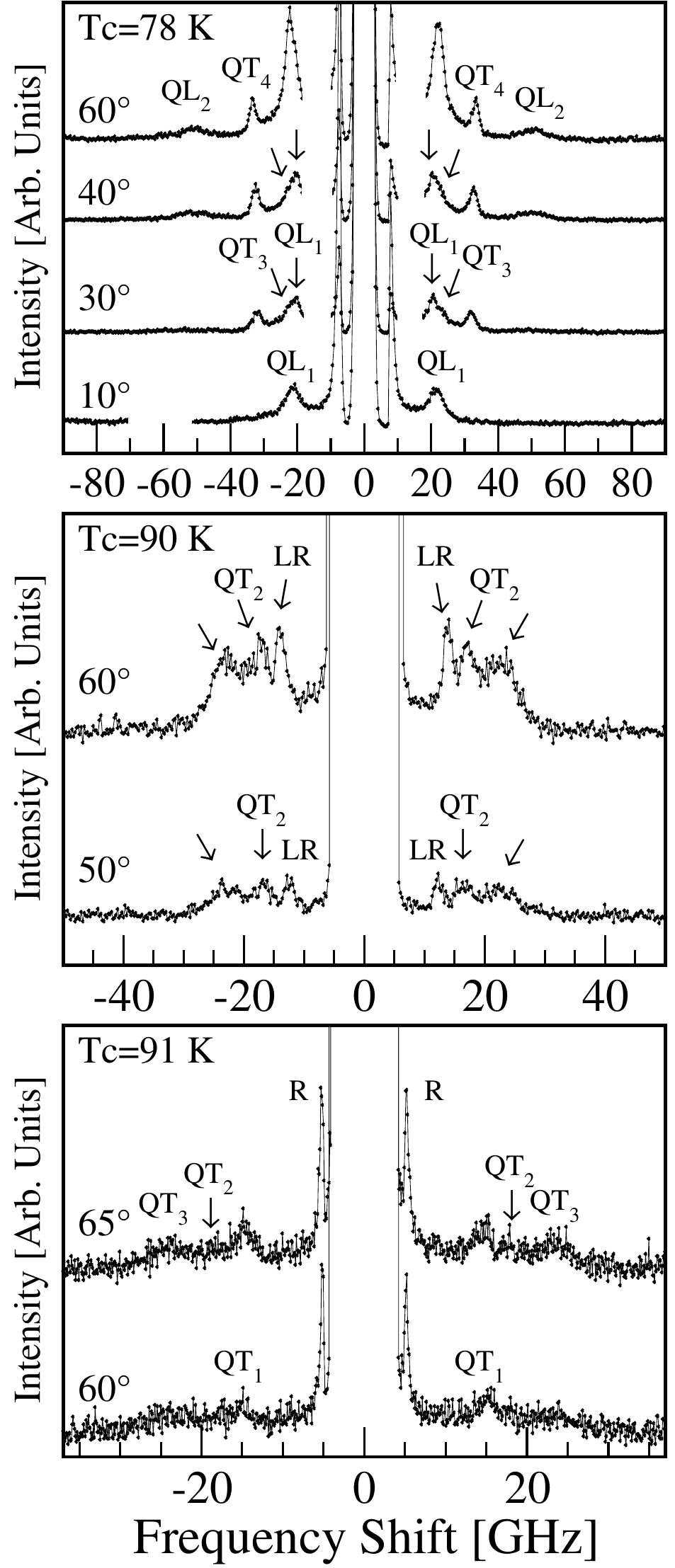} 
 \caption{Room-temperature Brillouin spectra of Bi-2212 single crystals with $T_c$ = 78 K (upper panel), $T_c$ = 90 K (middle panel), and $T_c$ = 91 K (lower panel), with eight acoustic phonon modes shown. R: surface Rayleigh, LR: Longitudinal surface resonance, QT: quasi-transverse, QL: quasi-longitudinal. Unlabelled arrows indicate a possible second mode emerging at higher incident angles. Note: In the top panel, $QT_1$ is omitted due it being overshadowed with sample tape. The region near $\sim60$ GHz is omitted in the 10$^\circ$ spectrum due to a known experimental artefact.}
 \label{fig:Spectra}
 \end{figure}

Fig.\;\ref{fig:Spectra} shows representative Brillouin spectra of three Bi-2212 crystals. Eight different Brillouin peak doublets (labelled $R$, $LR$, $QT_i$ with $i=1-4$, and $QL_j$ with $j=1-2$) are present in the spectra, with the number and intensity of these modes varying with angle of incidence and from sample to sample.  The overall quality of the spectra is strongly sample dependent with the $T_c=78$ K sample giving those of the highest quality.  In fact, spectra like those shown  in the top panel of Fig. \ref{fig:Spectra} were obtained on multiple flakes of this sample. Spectra obtained from the other two crystals are only of fair quality despite single-spectrum collection times of $\geq 20$ h. These differences may be due to crystal quality effects possibly related to growth method.  In any case, the Brillouin spectra in this work are of the highest quality yet obtained for Bi-2212. 

\begin{figure}[!t]
\includegraphics[width=\linewidth]{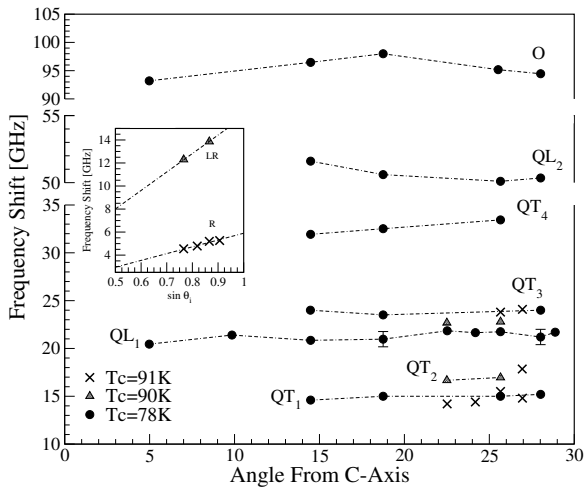}
\caption{Brillouin shift versus propagation direction for peaks due to bulk quasi-transverse ($QT_1$, $QT_2$, $QT_3$ $QT_4$), quasi-longitudinal ($QL_1$, $QL_2$) acoustic phonon modes, and a low-lying optic mode ($O$). The inset shows the Rayleigh ($R$) and longitudinal surface resonance ($LR$) acoustic phonon modes fitted through the origin. Omitted error bars are approximately the size of the associated data point symbols. }
\label{fig:Shift_vs_angle}
\end{figure}

The directional dependence of the Brillouin peak frequency shifts is shown in Fig. \ref{fig:Shift_vs_angle}.  With the exception of the $R$ and $LR$ peaks (see Fig. \ref{fig:Shift_vs_angle} inset), the shifts show only small variation with propagation direction over the range probed. For the lowest angles of incidence ({\it i.e.}, directions closest to the c-axis) only peak $QL_{1}$ is present in the spectra.  At larger angles of incidence ($> 14^{\circ}$ from the b-direction), the $R$, $LR$, $QT_{1}$, $QT_{2}$, $QT_{3}$, $QT_{4}$ and $QL_{2}$ peaks appear in spectra of one or more of the three samples.  It is noted that the large width and asymmetry of the peak at $\sim21$ GHz for the $T_c=90$ K sample (unlabelled peak indicated by arrows in middle panel of Fig. \ref{fig:Spectra}) suggests that it may in fact be two closely-spaced peaks due to the $QT_{2}$ and $QL_{1}$ modes, similar to what was observed for the $T_c=78$ K sample (see top panel of Fig. \ref{fig:Spectra}).   

\subsection{Mode Assignment}

\subsubsection{Surface Acoustic Phonon Modes}
The shifts of the peaks labelled $R$ and $LR$ show strong linear dependence on sine of the angle of incidence and are therefore assigned to the Rayleigh surface mode and longitudinal resonance, the latter being observed only in the spectra of the sample with $T_c = 90$ K.  Accordingly, the velocities of these modes were determined by fitting the Brillouin equation for surface modes, $f_S = 2V_S\sin{\theta_i}/\lambda_i$, where $V_S$ is the velocity of the Rayleigh mode ($S=R$) or the longitudinal resonance ($S=LR$), to the experimental frequency shift ($f_S$) versus $\sin{\theta_i}$ data.  The velocities, determined from the slopes of the lines of best-fit, were found to be $V_R = 1570$ m/s and $V_{LR} = 4260$ m/s for the $T_c = 91$ K and $T_c = 90$ K samples, respectively, and are comparable to those found in a previous Brillouin scattering study of Bi-2212 \cite{Boekholt}.  The widths of the $R$ and $LR$ peaks are also noticeably smaller than those of the other peaks, further supporting the assignment of these peaks to surface modes.

\subsubsection{Bulk Quasi-Transverse Acoustic Phonon Modes}
Peaks $QT_1$, $QT_2$, $QT_3$, and $QT_4$ are assigned to quasi-transverse acoustic modes because the shifts show little dependence on direction and because they are absent in small-incident-angle spectra \cite{Bottani2018} and show an overall increase in intensity with increasing angle of incidence.  Moreover, the frequency shifts of $QT_1$ and $QT_3$ are close to those of peaks identified as quasi-transverse acoustic modes in previous Brillouin scattering experiments \cite{Boekholt}, although it is also possible that the peak identified as the fast quasi-transverse mode in ref. \cite{Boekholt} is actually $QL_1$ due to its proximity to $QT_3$ as shown in the top panel of Fig. \ref{fig:Spectra}.  Further support for this mode assignment is provided by ultrasonics measurements on ceramic Bi-2212 samples with the $c$-axis of grains preferentially aligned which give transverse mode velocities within $\leq10$\% of those of $QT_1$ and $QT_2$ (see Table \ref{tab:BulkVelocities}).  Incidentally, the longitudinal acoustic mode velocity measured along the $c$-axis in the same study is also in excellent agreement with that of $QL_{1}$ determined in the present work \cite{Chang}.

It is also worth noting here that the close proximity of $QT_2$ and $QT_3$ to $QL_1$ (a range spanning $\sim6$ GHz) could explain why inelastic neutron scattering experiments have consistently measured only one transverse acoustic mode \cite{Etrillard2001}, since that technique lacks the resolution necessary to resolve two peaks separated by only a few GHz.



\begin{figure}[!t]
\includegraphics[scale=0.36]{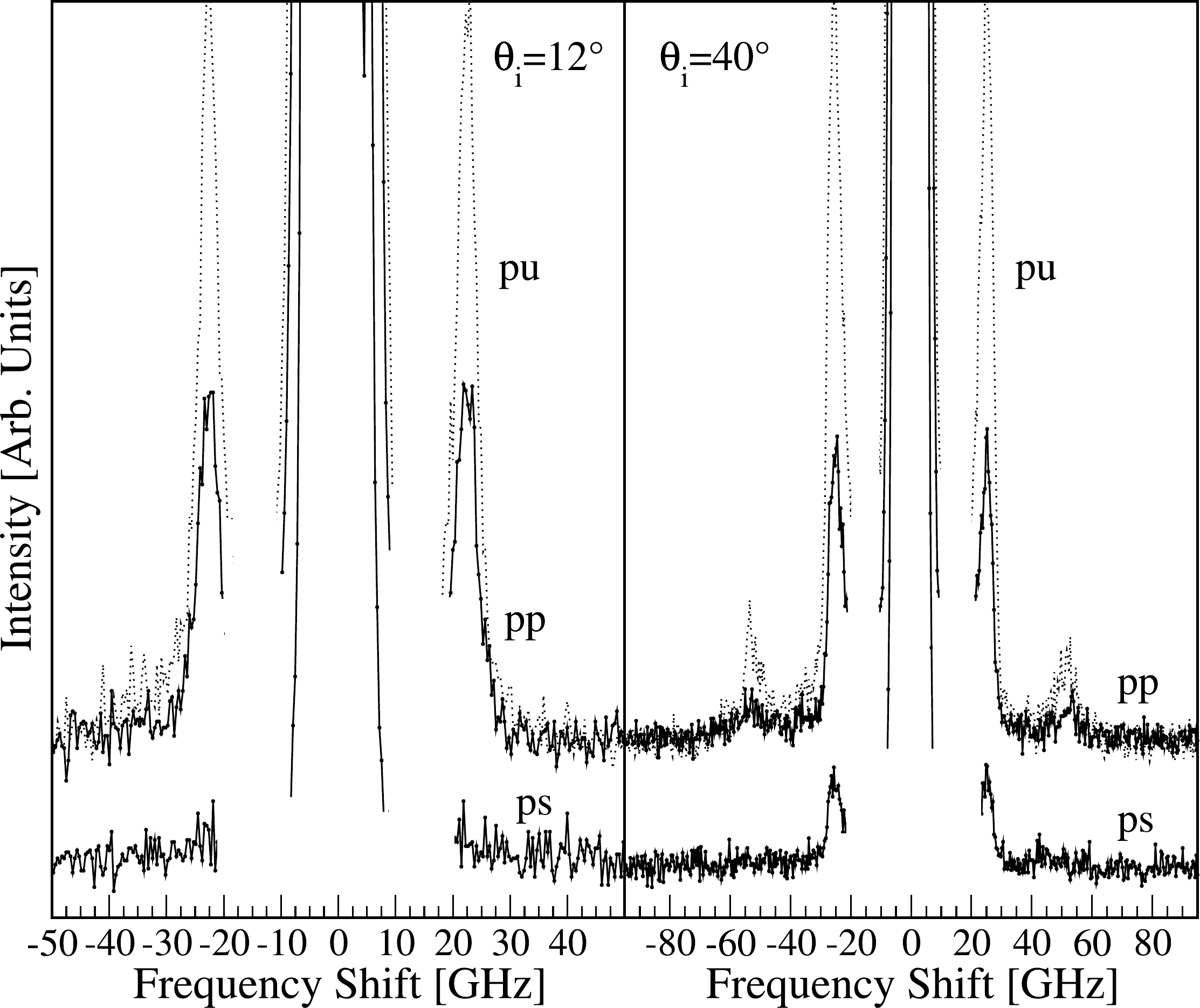} 
\caption{Polarized spectra of a Bi-2212 crystal with $T_c=78$ K.  The notation ``XY", where X, Y = u, p or s, indicates the polarization state of the incident and scattered light, respectively. u, p, and s refer to unpolarized, p-polarized, and s-polarized light, respectively. Note: The spectral region inside 20 GHz was removed because it contained a strong signal from sample mounting tape.}
\label{fig:Polarized_Spectra}
\end{figure}

\subsubsection{Bulk Quasi-Longitudinal Acoustic Phonon Modes}
The most surprising result of the present study is the presence of two distinct quasi-longitudinal bulk mode peaks ($QL_1$ and $QL_2$) in the Brillouin spectrum of Bi-2212, neither of which was seen in previous Brillouin scattering studies \cite{Boekholt, Baumgart,Aleksandrov}.  The longitudinal character of these modes was confirmed by polarization analysis of the scattered light.  To perform this analysis, it was first noted that for normally-incident p-polarized light, the light scattered from longitudinal and transverse bulk modes is p- and s-polarized, respectively \cite{cumm1972}.  Modes with longitudinal polarization will therefore be present in spectra for which the incident and scattered light are p-polarized (pp configuration) and absent when the incident light is p-polarized and the scattered light is s-polarized (ps configuration).  Although it was not possible to obtain spectra for normally-incident light, $QL_1$ was observed in pp-spectra obtained at an internal angle of incidence of $6^\circ$ to the crystallographic $c$-axis, but not in the corresponding ps-spectra, verifying that it is a longitudinal mode (see Fig. \ref{fig:Polarized_Spectra}).  A similar result was obtained for $QL_2$, but for an angle of $19^\circ$ from the $c$-axis because this mode could not be readily observed at lower angles (see Figs. \ref{fig:Spectra} and \ref{fig:Shift_vs_angle}).  It is noted that $QL_1$ is still visible but at reduced intensity in the $19^\circ$ ps-polarized spectrum due to the fact that this mode has some transverse character for such general directions of propagation. 

The fact that the $QL_1$ peak is quite intense in low incident angle spectra also supports its assignment to a quasi-longitudinal acoustic mode because strong peaks due to quasi-transverse modes are not expected to be present in Brillouin spectra under these circumstances \cite{Bottani2018}.  Furthermore, the rather large shift of the $QL_2$ peak ($\sim50$ GHz) is consistent with this assignment.

\subsubsection{Low-Lying Optic Phonon Mode} 
\begin{figure}[!t]
\centering
  \includegraphics[width=0.91\linewidth]{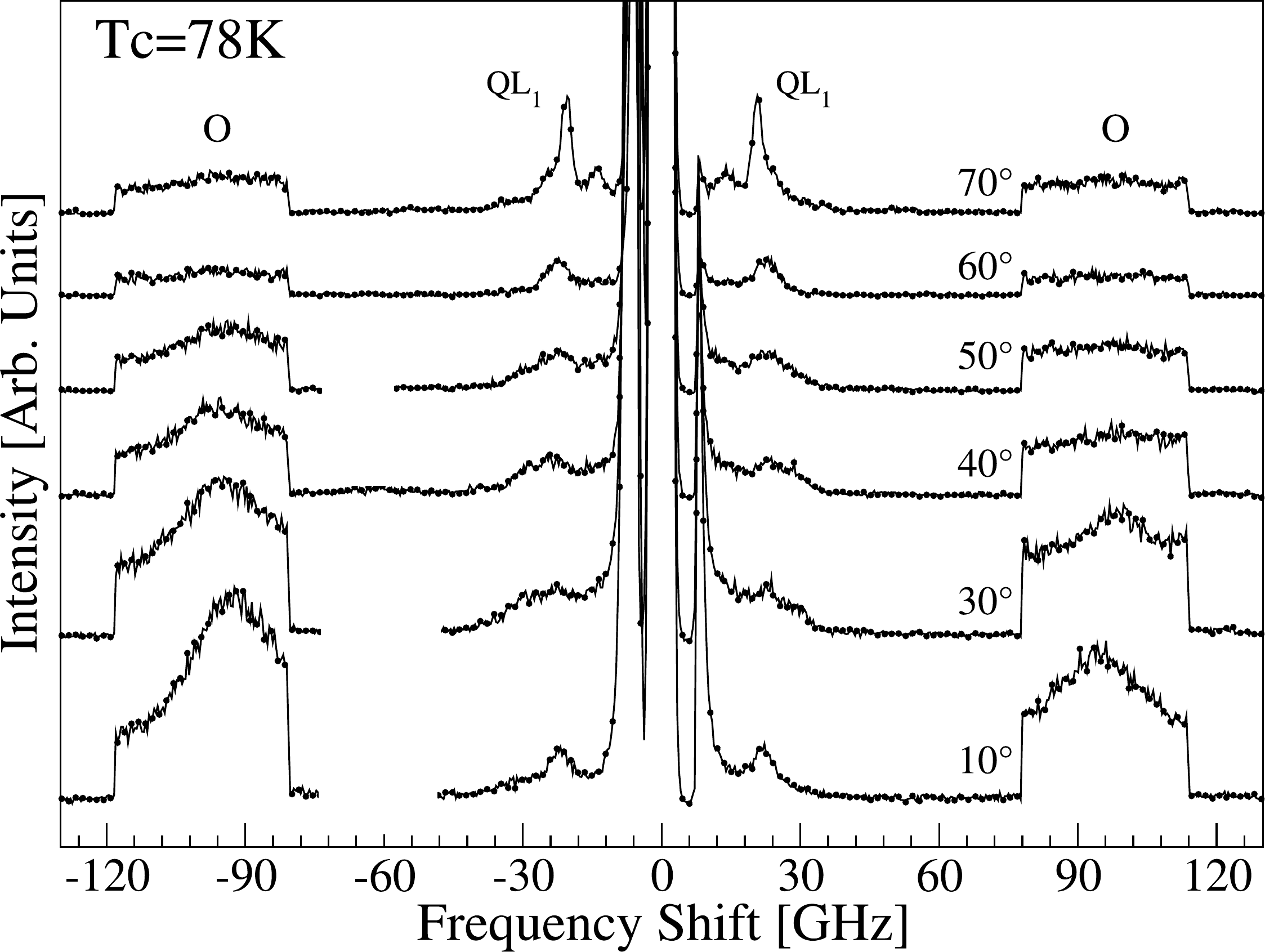}\\ 
   \includegraphics[width=\linewidth]{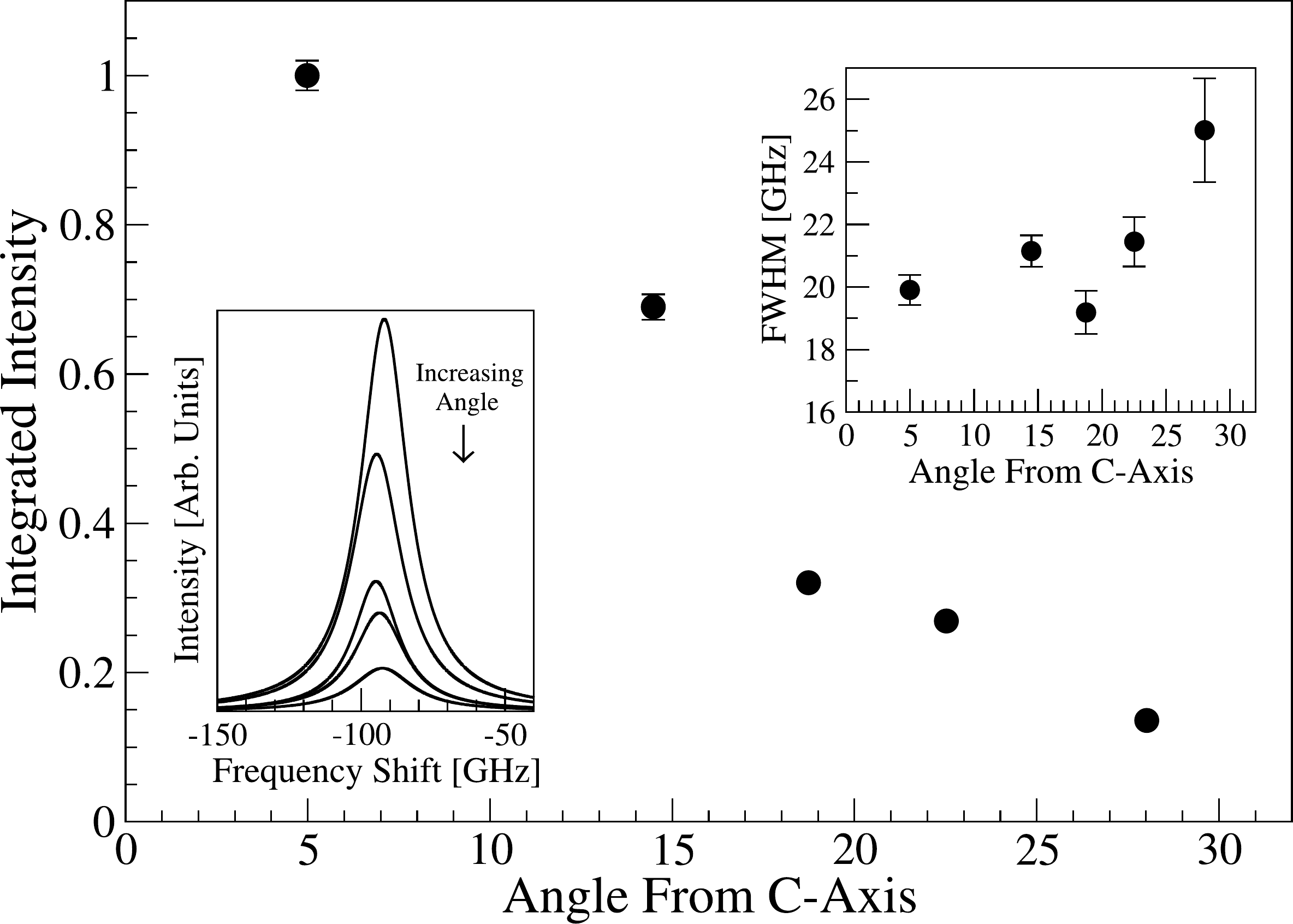}
 \caption{Top panel: Room-temperature Brillouin spectra of single crystal Bi-2212. The region from $\sim\pm80$ GHz to $\sim\pm120$ GHz was segmented such that the collection time per channel in this region was 10x that in the remainder of the spectrum.  Note: To avoid confusion, portions of some spectra are omitted due to a known artefact at $\sim60$ GHz. Bottom panel: Integrated intensity of a possible low-lying optic-like mode versus propagation direction as measured from the crystallographic $c$-axis. Left Inset: Peak intensity versus frequency shift for same mode.  Right Inset: FWHM of low-lying optic-like mode.} 
 \label{fig:95GHz_spectra}
 \end{figure}

Fig. \ref{fig:95GHz_spectra} shows that the Brillouin spectra contain a weak peak at $\sim\pm95$ GHz. The large frequency shift of this peak relative to others in the spectrum and the intensity difference between the Stokes and anti-Stokes scattering suggests that it is due to an extremely low-lying optic phonon mode. As the right panel of Fig. \ref{fig:95GHz_spectra} shows, this peak is strongest for phonon propagation directions close to the commensurate $c$-axis ($5^\circ$ from the $c$-axis for $\theta_i=10^\circ$) and becomes progressively less intense as the propagation direction moves away from this direction toward the $ab$-plane, being weakest at $\sim28^{\circ}$ ($\theta_i=70^\circ$) from the $c$-axis, the largest angle probed. This decrease in integrated intensity is accompanied by a corresponding decrease in peak intensity and an increase in peak width (FWHM), as seen in the left- and right-hand insets of Fig. \ref{fig:95GHz_spectra}, respectively. The peak frequency shift remains relatively constant over the range of directions probed, varying by a maximum of only 2-3\% from its mean value of 95.5 GHz (see Fig. \ref{fig:Shift_vs_angle}).  It was also noted that 
this peak appeared in both pp- and ps-spectra collected at angles of incidence very close to the $c$-axis ($<5^\circ$), suggesting that it is not due to a typical acoustic mode. 

\subsection{Incommensurate Structure}
The unusual Brillouin scattering results described above are a consequence of the incommensurate nature of Bi-2212. Surprisingly, this is the first time that the incommensurate character of Bi-2212 has been observed in the Brillouin spectrum - previous studies make no mention of its incommensurate structure nor are there any obvious signatures of it in published spectra.  This is primarily because the spectral region beyond shifts of $\pm30$ GHz was not explored, but may also be related to recent improvements in crystal quality. As seen in Fig.~\ref{fig:Spectra}, the presence of peaks in excess of those expected for a typical commensurate crystal is highly sample dependent and therefore may be a sensitive function of doping.  In fact, the presence or absence of some optic phonon peaks in Raman spectra of Bi-2212 and other cuprate systems has been shown to depend on doping \cite{Benhabib,Sulewski1991}.

\subsubsection{Signatures}
The most striking manifestation of the incommensurate structure of Bi-2212 in the Brillouin spectrum is the presence of two peaks, $QL_{1}$ and $QL_{2}$, due to quasi-longitudinal modes. These peaks are both present in spectra for phonon propagation directions between $\sim 15^{\circ}$ and $\sim 28^{\circ}$ from the $c$-axis.  In contrast, only $QL_1$ is present for propagation directions within $\sim 10^{\circ}$ of the commensurate $c$-direction (see  Fig.\ref{fig:Shift_vs_angle}).  This can be explained by noting that for phonon propagation directions at relatively large angles to the $c$-axis, the projection of the probed phonon wavevector in the $ab$ plane, and therefore the component of $\vec{q}$ along the incommensurate $b$-direction, will be appreciable, resulting in the presence of spectral features due to the incommensurate nature of Bi-2212, including $QL_2$.  As the direction of probed phonon propagation approaches the $c$-axis, the $b-$component of $\vec{q}$ approaches zero and the incommensurate nature is no longer apparent in the spectrum, accounting for the absence of $QL_2$ in small-angle spectra. The situation just described will be true unless, by coincidence, $\vec{q}$ happens to lie precisely in the $ac$ plane, in which case the $b$-component will be zero.  This is, however, extremely unlikely.

The presence of two quasi-longitudinal modes due to incommensurate structure in Brillouin spectra of Bi-2212 is consistent with the results of neutron scattering studies in which two longitudinal acoustic modes were observed along the incommensurate $b$-direction in Bi-2212 and Bi-2201 \cite{Etrillard2001,Etrillard2004}.  As Table \ref{tab:BulkVelocities} shows, the velocities of these modes for Bi-2212 were found to be 2400 m/s and 5900 m/s, and are comparable in magnitude to, and in the same ratio as, the velocities of the $QL_1$ and $QL_2$ modes (2720 m/s and 6700 m/s at 28$^{\circ}$ from c-axis, the closest direction to the $b$-axis probed; calculated using $V_B = f_B \lambda_i / 2n$, where $B$ refers to bulk and $n=2.0$ \cite{hwan2007}).  The fact that the velocities determined in the neutron scattering experiments are lower than those of $QL_1$ and $QL_2$ may seem counter-intuitive given that the magnitudes of the latter are partially determined by the elastic constant(s) associated with the weak interlayer bonding in Bi-2212. This can be reconciled by noting both the lower resolution of neutron scattering in the $q\rightarrow0$ limit when compared to Brillouin scattering and the fact that it probes acoustic phonons at much higher $q$ values where the velocity is typically expected to be lower due to a reduction in dispersion curve slope with increasing $q$.  This potential for the pertinent analytical techniques to measure different velocity values for a given mode due to changes in dispersion curve slope was also highlighted in Ref. \cite{Radulescu2002}. 

A second signature of the incommensurate structure of Bi-2212 is the existence of four distinct $QT$ modes for some propagation directions (see Fig. \ref{fig:Shift_vs_angle}). The observation of twice as many $QT$ peaks ({\it i.e.}, 2 sets of 2 $QT$ peaks) as would be expected for a commensurate crystal makes sense because of the two different periodicities associated with the incommensurate structure of Bi-2212.  This reasoning also nicely explains the presence of two $QL$ peaks in the Bi-2212 spectrum instead of the lone $QL$ peak normally observed in the spectrum of a commensurate crystal.

\begin{table*}[t]
\caption{Room temperature transverse (T) and longitudinal (L) bulk acoustic phonon velocities along the crystallographic axes for crystalline Bi-2212. Note: $V_T$/$V_L$ obtained from Brillouin scattering refers to quasi-transverse and quasi-longitudinal velocities, respectively.}
\begin{ruledtabular}
 \begin{tabular}{ c c c c c c c c c c  c}
\multirow{2}{*}{Technique} & \multirow{2}{*}{Study}& $T_c$ & \multirow{2}{*}{Direction} & $V_{T_1}$ & $V_{T_2}$ & $V_{T_3}$ & $V_{T_4}$ & $V_{L_1}$ & $V_{L_2}$ & $V_{L}$  \\
  & & [K] & & [m/s] & [m/s] & [m/s] & [m/s] & [m/s] & [m/s] & [m/s] \\
\hline
 \multirow{7}*{\begin{tabular}{l}
                   Brillouin\\
                   Light\\
                   Scattering\\
                   \end{tabular}} & \multirow{4}{*}{Present Work} & \multirow{2}{*}{78} & \multirow{1}{*}{[001]} & -- & -- & -- & -- & -- & -- & 2720\\
 & &  & 28$^\circ$ from [001] & 1930 & -- & 3190 & 4440 & 2720 & 6700 & 4710  \\
 & & 90 & 26$^\circ$ from [001] & -- & 2260 & -- & -- & -- & -- & --  \\
 & & 91 & 27$^\circ$ from [001] & 1968 & 2370 & -- & --  & --  & -- & -- \\
\cline{2-11}
 &  \multirow{2}{*}{Ref. \cite{Boekholt}\footnotemark[1]} &\multirow{2}{*}{78-92} & \multirow{1}{*}{[001]} & -- & -- & --  & --  & -- & -- & 3413   \\
&   &  & \multirow{1}{*}{[010]} & -- & -- & -- & -- & -- & --  & \multirow{1}{*}{4380}  \\
\cline{1-11}
 \multirow{3}*{\begin{tabular}{l}
                   Inelastic\\
                   Neutron\\
                   Scattering\\
                   \end{tabular}} & \multirow{2}{*}{Ref. \cite{Etrillard2001}} & \multirow{2}{*}{--} & \multirow{1}{*}{[001]} & 1830 & -- & -- & -- & --  & --    \\
 &  &  & \multirow{1}{*}{[010]} & -- & -- & -- & -- & 2400 & 5900 & 4200   \\         
 \cline{2-11}
  & \multirow{1}{*}{Ref. \cite{Merritt2019}\footnotemark[2]}  &  \multirow{1}{*}{--} & [001] & 1780 & -- & -- & -- & 2440 & -- & --\\
\cline{1-11}
\multirow{3}{*}{Ultrasonic} &  \multirow{1}{*}{Ref. \cite{Wang1,Wang2}\footnotemark[3]} &\multirow{1}{*}{84.5} & \multirow{1}{*}{[010]} & -- & 2460 & -- & -- &  -- & -- & 4150\\
 & \multirow{2}{*}{Ref. \cite{Chang}\footnotemark[4]}  & \multirow{2}{*}{--} & \multirow{1}{*}{[001]} & 1750 & -- & -- & -- & -- & -- & 2670 \\ 
& & & In (001) plane & 1740 & 2460 & -- & -- & -- & -- & 4370\\
\end{tabular}
\end{ruledtabular}
\label{tab:BulkVelocities}
\footnotetext[1]{Estimated from longitudinal resonance measurements, uncertainties are $\sim5\%$.}
\footnotetext[2]{Estimated from inelastic neutron scattering measurements at T$\leq$T$_c$.}
\footnotetext[3]{Estimated from velocity vs temperature curves.}
\footnotetext[4]{Velocities obtained from polycrystalline samples that show a preferred grain orientation.}
\end{table*}

The behaviour of the low-lying optic mode may also be a consequence of the incommensurate structure of Bi-2212. The trends in integrated intensity, peak intensity, FWHM, and peak frequency shift described above (see {\it Low-Lying Optic Phonon Mode} subsection) collectively suggest that the incommensurate structure of Bi-2212 causes this mode to be damped-out.  This reasoning may also explain why this mode is not observed in studies for which the primary or sole focus is phonon dynamics in the vicinity of the incommensurate $b$-direction.

\begin{figure}[!t]
\centering
  \includegraphics[width=\linewidth]{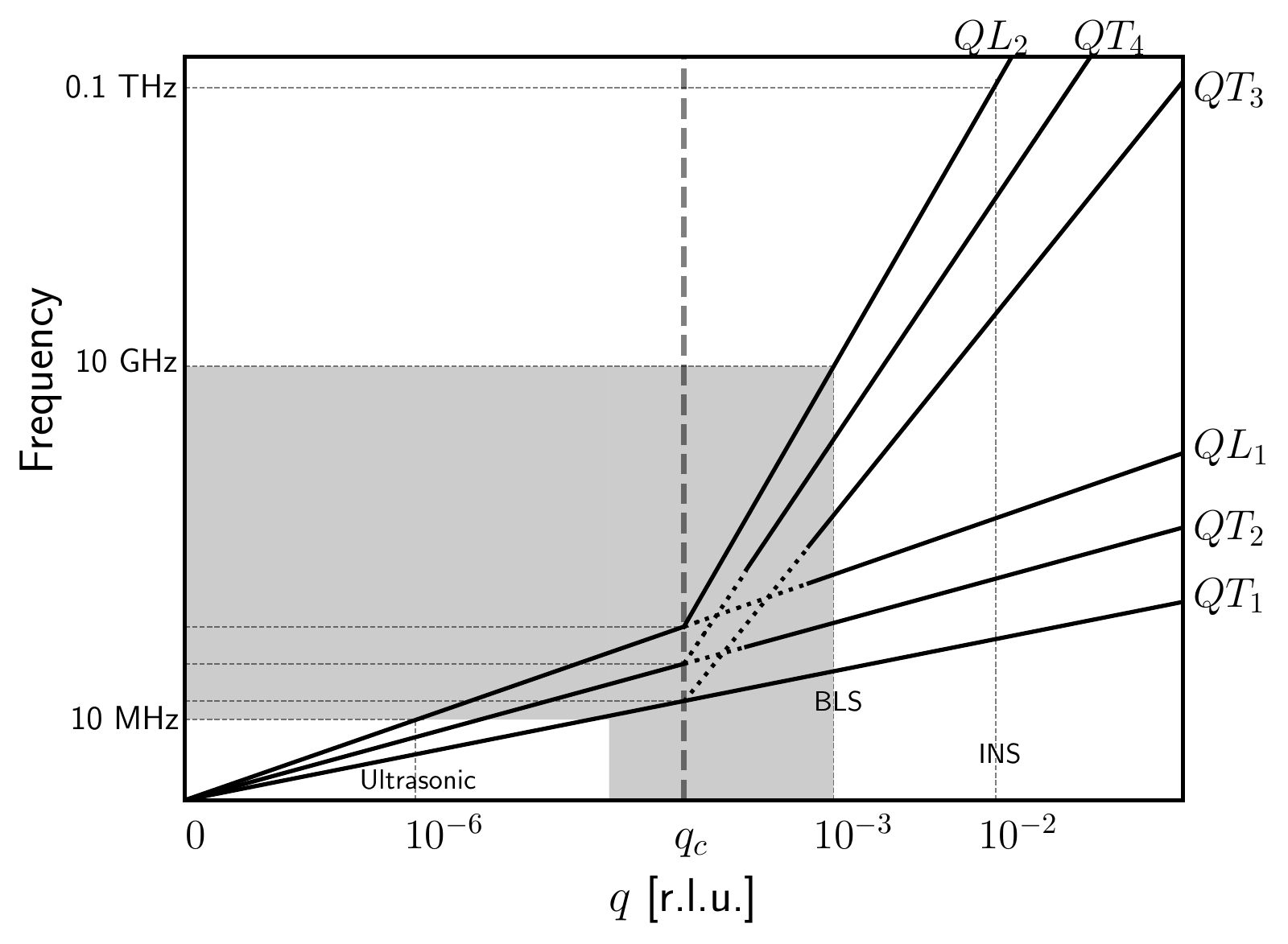}
 \caption{Schematic representation of low-$q$ acoustic phonon dispersion for Bi-2212 for a general direction of propagation. $q_c$ is the crossover frequency between commensurate and incommensurate dynamics. Approximate values of phonon wavevectors ($q$) in r.l.u. are shown for ultrasonic, Brillouin, and neutron scattering experiments.} 
 \label{fig:Schematic}
 \end{figure}

\subsubsection{Consistency with Composite Incommensurate Crystal Model}
The existence of two sets of three acoustic phonon modes (2 $\times$ (1 $QL$ + 2 $QT$)) suggests that Bi-2212 may be categorized as a composite incommensurate crystal, a class of aperiodic crystal in which the incommensurability arises from interpenetrating, mutually incommensurate sublattices \cite{radulescu1997} and for which, in contrast to modulated incommensurate crystals, an average periodicity cannot be defined \cite{Currat2002}.  This result also suggests that the two sublattices that comprise Bi-2212 are weakly interacting.  In fact, two independent sets of phonon branches are observed in a model consisting of two interpenetrating atomic chains of different periods when there is no coupling between the chains \cite{Currat2002}.  Moreover, while there do not appear to be any explicit theoretical or experimental results on acoustic phonon dispersion in composite incommensurate crystals for general phonon propagation directions, the results obtained in the present work do bear some similarities to the results of previous studies of acoustic phonon dynamics in high symmetry directions in Bi-2212. In particular, the presence of two distinct longitudinal acoustic modes is consistent with the predictions of the incommensurate composite model and also with inelastic neutron scattering results \cite{Etrillard2001, Etrillard2004}.  

Knowledge of the sublattice velocities $V_{QL_1}$ and $V_{QL_2}$ makes it possible to identify the composition of the two sublattices via their estimated mass ratio \cite{Finger1983}  
\begin{equation}
    \frac{m_1}{m_2}=\frac{V_{QL_{2}}^2-V_{q\rightarrow0}^2}{V_{q\rightarrow0}^2-V_{QL_{1}}^2},
    \label{eqn:massratio}
\end{equation}
where V$_{q\rightarrow0}$ velocity is the longitudinal mode velocity in the $q\rightarrow 0$ limit, and $m_i$ and $V_{QL_i}$, $i = 1,2$, are the mass and quasi-longitudinal velocity of the $i^{th}$ sublattice, respectively.
Given that ultrasonic experiments measure only a single longitudinal velocity equal to 4150 m/s in the MHz range in the incommensurate $b$-direction \cite{Wang1,Wang2}, this velocity can be taken as $V_{q\rightarrow0}$. Using this, along with $V_{QL_1}=2720$ m/s and $V_{QL_2}=6700$ m/s in Eq. \ref{eqn:massratio} gives $m_1/m_2=0.36$. The sublattices (neglecting $\delta$) Bi$_2$Sr$_2$O$_4$ with mass 658 u and CaCu$_2$O$_4$ with mass = 232 u, give $m_1/m_2 = 0.35$, in excellent agreement with the value obtained from Eq. \ref{eqn:massratio}.  This result is $\sim50$\% lower than the value $m_{1}/m_{2}=0.67$ obtained in neutron scattering studies which led to sublattice assignments of SrCaCu$_2$O$_6$ and Bi$_2$SrO$_{2+\delta}$. 

\subsubsection{Crossover Frequency}
Fig \ref{fig:Schematic} schematically shows the relatively low-$q$ portion of the acoustic phonon dispersion (low-lying optic mode not shown) for Bi-2212 for a general propagation direction as informed by collective consideration of ultrasonics \cite{Wang1,Wang2}, inelastic neutron scattering \cite{Etrillard2001}, and current Brillouin scattering data. At very small $q$, as probed by ultrasonic techniques at frequencies of $\sim10$ MHz, two transverse modes and a single longitudinal mode ($QT_1$, $QT_2$, and $QL_1$) would be observed as for a typical commensurate crystal. That this is the case can be inferred from the ``Ultrasonics" data at the bottom of Table \ref{tab:BulkVelocities}. At a critical wavevector, $q_c$, higher than has been probed in ultrasonics experiments, there is a crossover from three to six propagating bulk acoustic modes due to the incommensurate structure of Bi-2212. These six modes are those observed in the present Brillouin scattering study which probed phonons with a $q$-value greater than $q_c$ and with frequencies on the order of $\sim10$ GHz.  The frequency(ies) at which the crossover occurs therefore lies in the range 10 MHz - 10 GHz, one to four orders of magnitude higher than the value of $\sim 1$ MHz predicted by theory for the crossover from one to two propagating longitudinal acoustic modes \cite{Finger1983}.  Intuitively, one might expect an analogous crossover at comparable frequencies for quasi-transverse modes but this does not seem to be discussed in the literature.  Inelastic neutron scattering accesses phonons with still higher $q$ values in the $q>q_c$ region but with lower frequency resolution than Brillouin spectroscopy.  While there do not appear to be any measurements for directions other than along crystallographic axes, two longitudinal phonons were observed in the incommensurate $b$-direction in neutron scattering experiments on Bi-2212 \cite{Etrillard2001}, consistent with the scheme shown in Fig. \ref{fig:Schematic}. One of the consequences of the lower resolution, however, is that modes that are closely-spaced in frequency may not be resolved in neutron scattering experiments. This could explain the absence of otherwise anticipated modes in previous neutron scattering studies of Bi-2212 \cite{Etrillard2001}.

\section{Conclusion}
In summary, low-frequency phonon dynamics of Bi-2212 single crystals were studied by Brillouin light scattering spectroscopy. From collected spectra, two quasi-longitudinal acoustic and four quasi-transverse acoustic phonon modes were observed, with the former in agreement with previous inelastic neutron scattering studies \cite{Etrillard2001} indicating that Bi-2212 is a composite incommensurate crystal. The measured frequency shifts of the six bulk acoustic phonon modes allowed for the assignment of each mode to a given sublattice, while the quasi-longitudinal acoustic phonon velocities were used to propose sublattice assignments of Bi$_2$Sr$_2$O$_4$ and CaCu$_2$O$_4$. Furthermore, a low-lying optic-like mode was also observed and appears to be another manifestation of the incommensurate nature of Bi-2212. The rich and highly unusual Brillouin spectra obtained here make it clear that further attention is warranted to fully understand phonon behaviour in Bi-2212 in the $q\rightarrow0$ limit. Moreover, the insights into long-wavelength phonon dynamics revealed in the current work will lead to a deeper understanding of the role of phonons and of electron-phonon coupling in high-T$_c$ superconductivity.

\acknowledgements
The authors would like to acknowledge Dr. J. P. Clancy at McMaster University, Canada for supplying the samples used in this work. 
\bibliographystyle{apsrev4-2}
\bibliography{refs.bib}

\begin{thebibliography}{25}%
\makeatletter
\providecommand \@ifxundefined [1]{%
 \@ifx{#1\undefined}
}%
\providecommand \@ifnum [1]{%
 \ifnum #1\expandafter \@firstoftwo
 \else \expandafter \@secondoftwo
 \fi
}%
\providecommand \@ifx [1]{%
 \ifx #1\expandafter \@firstoftwo
 \else \expandafter \@secondoftwo
 \fi
}%
\providecommand \natexlab [1]{#1}%
\providecommand \enquote  [1]{``#1''}%
\providecommand \bibnamefont  [1]{#1}%
\providecommand \bibfnamefont [1]{#1}%
\providecommand \citenamefont [1]{#1}%
\providecommand \href@noop [0]{\@secondoftwo}%
\providecommand \href [0]{\begingroup \@sanitize@url \@href}%
\providecommand \@href[1]{\@@startlink{#1}\@@href}%
\providecommand \@@href[1]{\endgroup#1\@@endlink}%
\providecommand \@sanitize@url [0]{\catcode `\\12\catcode `\$12\catcode
  `\&12\catcode `\#12\catcode `\^12\catcode `\_12\catcode `\%12\relax}%
\providecommand \@@startlink[1]{}%
\providecommand \@@endlink[0]{}%
\providecommand \url  [0]{\begingroup\@sanitize@url \@url }%
\providecommand \@url [1]{\endgroup\@href {#1}{\urlprefix }}%
\providecommand \urlprefix  [0]{URL }%
\providecommand \Eprint [0]{\href }%
\providecommand \doibase [0]{https://doi.org/}%
\providecommand \selectlanguage [0]{\@gobble}%
\providecommand \bibinfo  [0]{\@secondoftwo}%
\providecommand \bibfield  [0]{\@secondoftwo}%
\providecommand \translation [1]{[#1]}%
\providecommand \BibitemOpen [0]{}%
\providecommand \bibitemStop [0]{}%
\providecommand \bibitemNoStop [0]{.\EOS\space}%
\providecommand \EOS [0]{\spacefactor3000\relax}%
\providecommand \BibitemShut  [1]{\csname bibitem#1\endcsname}%
\let\auto@bib@innerbib\@empty
\bibitem [{\citenamefont {Axe}\ and\ \citenamefont {Bak}(1982)}]{Axe1982}%
  \BibitemOpen
  \bibfield  {author} {\bibinfo {author} {\bibfnamefont {J.~D.}\ \bibnamefont
  {Axe}}\ and\ \bibinfo {author} {\bibfnamefont {P.}~\bibnamefont {Bak}},\
  }\href {https://doi.org/10.1103/PhysRevB.26.4963} {\bibfield  {journal}
  {\bibinfo  {journal} {Phys. Rev. B}\ }\textbf {\bibinfo {volume} {26}},\
  \bibinfo {pages} {4963} (\bibinfo {year} {1982})}\BibitemShut {NoStop}%
\bibitem [{\citenamefont {Currat}\ \emph {et~al.}(2002)\citenamefont {Currat},
  \citenamefont {Kats},\ and\ \citenamefont {Luk'yanchuk}}]{Currat2002}%
  \BibitemOpen
  \bibfield  {author} {\bibinfo {author} {\bibfnamefont {R.}~\bibnamefont
  {Currat}}, \bibinfo {author} {\bibfnamefont {E.}~\bibnamefont {Kats}},\ and\
  \bibinfo {author} {\bibfnamefont {I.}~\bibnamefont {Luk'yanchuk}},\ }\href
  {https://doi.org/10.1140/epjb/e20020098} {\bibfield  {journal} {\bibinfo
  {journal} {Eur. Phys. J. B}\ }\textbf {\bibinfo {volume} {26}},\ \bibinfo
  {pages} {339} (\bibinfo {year} {2002})}\BibitemShut {NoStop}%
\bibitem [{\citenamefont {Dzugutov}\ and\ \citenamefont
  {Phillips}(1995)}]{Dzugutov1995}%
  \BibitemOpen
  \bibfield  {author} {\bibinfo {author} {\bibfnamefont {M.}~\bibnamefont
  {Dzugutov}}\ and\ \bibinfo {author} {\bibfnamefont {J.}~\bibnamefont
  {Phillips}},\ }\href
  {https://doi.org/https://doi.org/10.1016/0022-3093(95)00425-4} {\bibfield
  {journal} {\bibinfo  {journal} {J Non-Cryst. Solids}\ }\textbf {\bibinfo
  {volume} {192-193}},\ \bibinfo {pages} {397} (\bibinfo {year}
  {1995})}\BibitemShut {NoStop}%
\bibitem [{\citenamefont {Finger}\ and\ \citenamefont
  {Rice}(1982)}]{Finger1982}%
  \BibitemOpen
  \bibfield  {author} {\bibinfo {author} {\bibfnamefont {W.}~\bibnamefont
  {Finger}}\ and\ \bibinfo {author} {\bibfnamefont {T.~M.}\ \bibnamefont
  {Rice}},\ }\href {https://doi.org/10.1103/PhysRevLett.49.468} {\bibfield
  {journal} {\bibinfo  {journal} {Phys. Rev. Lett.}\ }\textbf {\bibinfo
  {volume} {49}},\ \bibinfo {pages} {468} (\bibinfo {year} {1982})}\BibitemShut
  {NoStop}%
\bibitem [{\citenamefont {Janssen}\ \emph {et~al.}(2002)\citenamefont
  {Janssen}, \citenamefont {Radulescu},\ and\ \citenamefont
  {Rubtsov}}]{Janssen2002}%
  \BibitemOpen
  \bibfield  {author} {\bibinfo {author} {\bibfnamefont {T.}~\bibnamefont
  {Janssen}}, \bibinfo {author} {\bibfnamefont {O.}~\bibnamefont {Radulescu}},\
  and\ \bibinfo {author} {\bibfnamefont {A.~N.}\ \bibnamefont {Rubtsov}},\
  }\href {https://doi.org/10.1140/epjb/e2002-00265-y} {\bibfield  {journal}
  {\bibinfo  {journal} {Eur. Phys. J. B}\ }\textbf {\bibinfo {volume} {29}},\
  \bibinfo {pages} {85} (\bibinfo {year} {2002})}\BibitemShut {NoStop}%
\bibitem [{\citenamefont {{O. Radulescu, T. Janssen and J.
  Etrillard}}(2002)}]{Radulescu2002}%
  \BibitemOpen
  \bibfield  {author} {\bibinfo {author} {\bibnamefont {{O. Radulescu, T.
  Janssen and J. Etrillard}}},\ }\href
  {https://doi.org/10.1140/epjb/e2002-00322-7} {\bibfield  {journal} {\bibinfo
  {journal} {Eur. Phys. J. B}\ }\textbf {\bibinfo {volume} {29}},\ \bibinfo
  {pages} {385} (\bibinfo {year} {2002})}\BibitemShut {NoStop}%
\bibitem [{\citenamefont {Miles}\ \emph {et~al.}(1998)\citenamefont {Miles},
  \citenamefont {Kennedy}, \citenamefont {McIntyre}, \citenamefont {Gu},
  \citenamefont {Russell},\ and\ \citenamefont {Koshizuka}}]{Miles1998}%
  \BibitemOpen
  \bibfield  {author} {\bibinfo {author} {\bibfnamefont {P.}~\bibnamefont
  {Miles}}, \bibinfo {author} {\bibfnamefont {S.}~\bibnamefont {Kennedy}},
  \bibinfo {author} {\bibfnamefont {G.}~\bibnamefont {McIntyre}}, \bibinfo
  {author} {\bibfnamefont {G.}~\bibnamefont {Gu}}, \bibinfo {author}
  {\bibfnamefont {G.}~\bibnamefont {Russell}},\ and\ \bibinfo {author}
  {\bibfnamefont {N.}~\bibnamefont {Koshizuka}},\ }\href
  {https://doi.org/https://doi.org/10.1016/S0921-4534(97)01682-1} {\bibfield
  {journal} {\bibinfo  {journal} {Physica C Supercond.}\ }\textbf {\bibinfo
  {volume} {294}},\ \bibinfo {pages} {275} (\bibinfo {year}
  {1998})}\BibitemShut {NoStop}%
\bibitem [{\citenamefont {Kan}\ and\ \citenamefont {Moss}(1992)}]{Kan1992}%
  \BibitemOpen
  \bibfield  {author} {\bibinfo {author} {\bibfnamefont {X.~B.}\ \bibnamefont
  {Kan}}\ and\ \bibinfo {author} {\bibfnamefont {S.~C.}\ \bibnamefont {Moss}},\
  }\href {https://doi.org/10.1107/S0108768191011333} {\bibfield  {journal}
  {\bibinfo  {journal} {Acta Crystallographica Section B}\ }\textbf {\bibinfo
  {volume} {48}},\ \bibinfo {pages} {122} (\bibinfo {year} {1992})}\BibitemShut
  {NoStop}%
\bibitem [{\citenamefont {Ariosa}\ \emph {et~al.}(2001)\citenamefont {Ariosa},
  \citenamefont {Berger}, \citenamefont {Schmauder}, \citenamefont {Pavuna},
  \citenamefont {Margaritondo}, \citenamefont {Christensen}, \citenamefont
  {Kelley},\ and\ \citenamefont {Onellion}}]{Ariosa2001}%
  \BibitemOpen
  \bibfield  {author} {\bibinfo {author} {\bibfnamefont {D.}~\bibnamefont
  {Ariosa}}, \bibinfo {author} {\bibfnamefont {H.}~\bibnamefont {Berger}},
  \bibinfo {author} {\bibfnamefont {T.}~\bibnamefont {Schmauder}}, \bibinfo
  {author} {\bibfnamefont {D.}~\bibnamefont {Pavuna}}, \bibinfo {author}
  {\bibfnamefont {G.}~\bibnamefont {Margaritondo}}, \bibinfo {author}
  {\bibfnamefont {S.}~\bibnamefont {Christensen}}, \bibinfo {author}
  {\bibfnamefont {R.}~\bibnamefont {Kelley}},\ and\ \bibinfo {author}
  {\bibfnamefont {M.}~\bibnamefont {Onellion}},\ }\href
  {https://doi.org/https://doi.org/10.1016/S0921-4534(00)01629-4} {\bibfield
  {journal} {\bibinfo  {journal} {Physica C Supercond.}\ }\textbf {\bibinfo
  {volume} {351}},\ \bibinfo {pages} {251} (\bibinfo {year}
  {2001})}\BibitemShut {NoStop}%
\bibitem [{\citenamefont {{A. M. Merritt, J.-P. Castellan, T. Keller, S.R.
  Park, J.A. Fernandez-Baca, G.D. Gu and D. Reznik}}(2019)}]{Merritt2019}%
  \BibitemOpen
  \bibfield  {author} {\bibinfo {author} {\bibnamefont {{A. M. Merritt, J.-P.
  Castellan, T. Keller, S.R. Park, J.A. Fernandez-Baca, G.D. Gu and D.
  Reznik}}},\ }\href {https://doi.org/10.1103/PhysRevB.100.144502} {\bibfield
  {journal} {\bibinfo  {journal} {Phys. Rev. B}\ }\textbf {\bibinfo {volume}
  {100}},\ \bibinfo {pages} {144502} (\bibinfo {year} {2019})}\BibitemShut
  {NoStop}%
\bibitem [{\citenamefont {{J. Etrillard, Ph Bourges, H.F. He, B. Keimer, B.
  Liang and C.T. Lin}}(2001)}]{Etrillard2001}%
  \BibitemOpen
  \bibfield  {author} {\bibinfo {author} {\bibnamefont {{J. Etrillard, Ph
  Bourges, H.F. He, B. Keimer, B. Liang and C.T. Lin}}},\ }\href
  {https://doi.org/10.1209/epl/i2001-00400-0} {\bibfield  {journal} {\bibinfo
  {journal} {Europhys. Lett.}\ }\textbf {\bibinfo {volume} {55}},\ \bibinfo
  {pages} {201} (\bibinfo {year} {2001})}\BibitemShut {NoStop}%
\bibitem [{\citenamefont {{Y. Wang, J. Wu, J. Zhu, H. Shen, Y. Yan and Z.
  Zhao}}(1989)}]{Wang1}%
  \BibitemOpen
  \bibfield  {author} {\bibinfo {author} {\bibnamefont {{Y. Wang, J. Wu, J.
  Zhu, H. Shen, Y. Yan and Z. Zhao}}},\ }\href
  {https://doi.org/https://doi.org/10.1016/0921-4534(89)91102-7} {\bibfield
  {journal} {\bibinfo  {journal} {Physica C}\ }\textbf {\bibinfo {volume}
  {162-164}},\ \bibinfo {pages} {454} (\bibinfo {year} {1989})}\BibitemShut
  {NoStop}%
\bibitem [{\citenamefont {{J. Etrillard, L. Bourges, B. Liang, C.T. Lin and B.
  Keimer}}(2004)}]{Etrillard2004}%
  \BibitemOpen
  \bibfield  {author} {\bibinfo {author} {\bibnamefont {{J. Etrillard, L.
  Bourges, B. Liang, C.T. Lin and B. Keimer}}},\ }\href
  {https://doi.org/10.1209/epl/i2003-10182-3} {\bibfield  {journal} {\bibinfo
  {journal} {Europhys. Lett.}\ }\textbf {\bibinfo {volume} {66}},\ \bibinfo
  {pages} {246} (\bibinfo {year} {2004})}\BibitemShut {NoStop}%
\bibitem [{\citenamefont {{M. Boekholt, J.V. Harzer, B. Hillebrands and G.
  G{\"u}ntherodt}}(1991)}]{Boekholt}%
  \BibitemOpen
  \bibfield  {author} {\bibinfo {author} {\bibnamefont {{M. Boekholt, J.V.
  Harzer, B. Hillebrands and G. G{\"u}ntherodt}}},\ }\href
  {https://doi.org/https://doi.org/10.1016/0921-4534(91)90017-S} {\bibfield
  {journal} {\bibinfo  {journal} {Physica C}\ }\textbf {\bibinfo {volume}
  {179}},\ \bibinfo {pages} {101} (\bibinfo {year} {1991})}\BibitemShut
  {NoStop}%
\bibitem [{\citenamefont {Bottani}\ and\ \citenamefont
  {Fioretto}(2018)}]{Bottani2018}%
  \BibitemOpen
  \bibfield  {author} {\bibinfo {author} {\bibfnamefont {C.~E.}\ \bibnamefont
  {Bottani}}\ and\ \bibinfo {author} {\bibfnamefont {D.}~\bibnamefont
  {Fioretto}},\ }\href@noop {} {\bibfield  {journal} {\bibinfo  {journal} {Adv.
  Phys. X}\ }\textbf {\bibinfo {volume} {3}},\ \bibinfo {pages} {607} (\bibinfo
  {year} {2018})}\BibitemShut {NoStop}%
\bibitem [{\citenamefont {Chang}\ \emph {et~al.}(1993)\citenamefont {Chang},
  \citenamefont {Ford}, \citenamefont {Saunders}, \citenamefont {Jiaqiang},
  \citenamefont {Almond}, \citenamefont {Chapman}, \citenamefont {Cankurtaran},
  \citenamefont {Poeppel},\ and\ \citenamefont {Goretta}}]{Chang}%
  \BibitemOpen
  \bibfield  {author} {\bibinfo {author} {\bibfnamefont {F.}~\bibnamefont
  {Chang}}, \bibinfo {author} {\bibfnamefont {P.~J.}\ \bibnamefont {Ford}},
  \bibinfo {author} {\bibfnamefont {G.~A.}\ \bibnamefont {Saunders}}, \bibinfo
  {author} {\bibfnamefont {L.}~\bibnamefont {Jiaqiang}}, \bibinfo {author}
  {\bibfnamefont {D.~P.}\ \bibnamefont {Almond}}, \bibinfo {author}
  {\bibfnamefont {B.}~\bibnamefont {Chapman}}, \bibinfo {author} {\bibfnamefont
  {M.}~\bibnamefont {Cankurtaran}}, \bibinfo {author} {\bibfnamefont {R.~B.}\
  \bibnamefont {Poeppel}},\ and\ \bibinfo {author} {\bibfnamefont {K.~C.}\
  \bibnamefont {Goretta}},\ }\href {https://doi.org/10.1088/0953-2048/6/7/006}
  {\bibfield  {journal} {\bibinfo  {journal} {Supercond. Sci. Technol.}\
  }\textbf {\bibinfo {volume} {6}},\ \bibinfo {pages} {484} (\bibinfo {year}
  {1993})}\BibitemShut {NoStop}%
\bibitem [{\citenamefont {{P. Baumgart, S. Blumenr{\"o}der, A. Erle, B.
  Hillebrands, P. Splittgerber, G. G{\"u}ntherodt and H.
  Schmidt}}(1989)}]{Baumgart}%
  \BibitemOpen
  \bibfield  {author} {\bibinfo {author} {\bibnamefont {{P. Baumgart, S.
  Blumenr{\"o}der, A. Erle, B. Hillebrands, P. Splittgerber, G. G{\"u}ntherodt
  and H. Schmidt}}},\ }\href
  {https://doi.org/https://doi.org/10.1016/0921-4534(89)90599-6} {\bibfield
  {journal} {\bibinfo  {journal} {Physica C}\ }\textbf {\bibinfo {volume}
  {162-164}},\ \bibinfo {pages} {1073} (\bibinfo {year} {1989})}\BibitemShut
  {NoStop}%
\bibitem [{\citenamefont {{V. V. Aleksandrov, T. S. Velichkina, V. I.
  Voronkova, A. A. Gippius, S. V. Rek'ko, I. A. Yakovlev and V. K.
  Yanovskii}}(1990)}]{Aleksandrov}%
  \BibitemOpen
  \bibfield  {author} {\bibinfo {author} {\bibnamefont {{V. V. Aleksandrov, T.
  S. Velichkina, V. I. Voronkova, A. A. Gippius, S. V. Rek'ko, I. A. Yakovlev
  and V. K. Yanovskii}}},\ }\href
  {https://doi.org/https://doi.org/10.1016/0038-1098(90)90116-S} {\bibfield
  {journal} {\bibinfo  {journal} {Solid State Commun.}\ }\textbf {\bibinfo
  {volume} {76}},\ \bibinfo {pages} {685} (\bibinfo {year} {1990})}\BibitemShut
  {NoStop}%
\bibitem [{\citenamefont {Cummins}\ and\ \citenamefont
  {Schoen}(1972)}]{cumm1972}%
  \BibitemOpen
  \bibfield  {author} {\bibinfo {author} {\bibfnamefont {H.~Z.}\ \bibnamefont
  {Cummins}}\ and\ \bibinfo {author} {\bibfnamefont {P.~E.}\ \bibnamefont
  {Schoen}},\ }\bibinfo {title} {Linear scattering from thermal fluctuations},\
  in\ \href@noop {} {\emph {\bibinfo {booktitle} {Laser Handbook}}},\
  Vol.~\bibinfo {volume} {2},\ \bibinfo {editor} {edited by\ \bibinfo {editor}
  {\bibfnamefont {F.~T.}\ \bibnamefont {Arecchi}}, \bibinfo {editor}
  {\bibfnamefont {E.~O.}\ \bibnamefont {Schulz-Dubois}},\ and\ \bibinfo
  {editor} {\bibfnamefont {M.~L.}\ \bibnamefont {Stitch}}}\ (\bibinfo
  {publisher} {North-Holland},\ \bibinfo {address} {New York},\ \bibinfo {year}
  {1972})\ Chap.~\bibinfo {chapter} {E1}, pp.\ \bibinfo {pages}
  {1029--1075}\BibitemShut {NoStop}%
\bibitem [{\citenamefont {Benhabib}\ \emph {et~al.}(2015)\citenamefont
  {Benhabib}, \citenamefont {Gallais}, \citenamefont {Cazayous}, \citenamefont
  {M\'easson}, \citenamefont {Zhong}, \citenamefont {Schneeloch}, \citenamefont
  {Forget}, \citenamefont {Gu}, \citenamefont {Colson},\ and\ \citenamefont
  {Sacuto}}]{Benhabib}%
  \BibitemOpen
  \bibfield  {author} {\bibinfo {author} {\bibfnamefont {S.}~\bibnamefont
  {Benhabib}}, \bibinfo {author} {\bibfnamefont {Y.}~\bibnamefont {Gallais}},
  \bibinfo {author} {\bibfnamefont {M.}~\bibnamefont {Cazayous}}, \bibinfo
  {author} {\bibfnamefont {M.-A.}\ \bibnamefont {M\'easson}}, \bibinfo {author}
  {\bibfnamefont {R.~D.}\ \bibnamefont {Zhong}}, \bibinfo {author}
  {\bibfnamefont {J.}~\bibnamefont {Schneeloch}}, \bibinfo {author}
  {\bibfnamefont {A.}~\bibnamefont {Forget}}, \bibinfo {author} {\bibfnamefont
  {G.~D.}\ \bibnamefont {Gu}}, \bibinfo {author} {\bibfnamefont
  {D.}~\bibnamefont {Colson}},\ and\ \bibinfo {author} {\bibfnamefont
  {A.}~\bibnamefont {Sacuto}},\ }\href
  {https://doi.org/10.1103/PhysRevB.92.134502} {\bibfield  {journal} {\bibinfo
  {journal} {Phys. Rev. B}\ }\textbf {\bibinfo {volume} {92}},\ \bibinfo
  {pages} {134502} (\bibinfo {year} {2015})}\BibitemShut {NoStop}%
\bibitem [{\citenamefont {Sulewski}\ \emph {et~al.}(1991)\citenamefont
  {Sulewski}, \citenamefont {Fleury},\ and\ \citenamefont
  {Lyons}}]{Sulewski1991}%
  \BibitemOpen
  \bibfield  {author} {\bibinfo {author} {\bibfnamefont {P.~E.}\ \bibnamefont
  {Sulewski}}, \bibinfo {author} {\bibfnamefont {P.~A.}\ \bibnamefont
  {Fleury}},\ and\ \bibinfo {author} {\bibfnamefont {K.~B.}\ \bibnamefont
  {Lyons}},\ }\bibinfo {title} {Light scattering in oxide superconductors},\
  in\ \href@noop {} {\emph {\bibinfo {booktitle} {Laser Optics of Condensed
  Matter: Volume 2 The Physics of Optical Phenomena and Their Use as Probes of
  Matter}}},\ \bibinfo {editor} {edited by\ \bibinfo {editor} {\bibfnamefont
  {E.}~\bibnamefont {Garmire}}, \bibinfo {editor} {\bibfnamefont {A.~A.}\
  \bibnamefont {Maradudin}},\ and\ \bibinfo {editor} {\bibfnamefont {K.~K.}\
  \bibnamefont {Rebane}}}\ (\bibinfo  {publisher} {Springer US},\ \bibinfo
  {address} {Boston, MA},\ \bibinfo {year} {1991})\ pp.\ \bibinfo {pages}
  {121--130}\BibitemShut {NoStop}%
\bibitem [{\citenamefont {Hwang}\ \emph {et~al.}(2007)\citenamefont {Hwang},
  \citenamefont {Timusk},\ and\ \citenamefont {Gu}}]{hwan2007}%
  \BibitemOpen
  \bibfield  {author} {\bibinfo {author} {\bibfnamefont {J.}~\bibnamefont
  {Hwang}}, \bibinfo {author} {\bibfnamefont {T.}~\bibnamefont {Timusk}},\ and\
  \bibinfo {author} {\bibfnamefont {G.}~\bibnamefont {Gu}},\ }\href
  {https://doi.org/10.1088/0953-8984/19/12/125208} {\bibfield  {journal}
  {\bibinfo  {journal} {J. Phys. Condens. Matter}\ }\textbf {\bibinfo {volume}
  {19}},\ \bibinfo {pages} {125208} (\bibinfo {year} {2007})}\BibitemShut
  {NoStop}%
\bibitem [{\citenamefont {Wang}\ \emph {et~al.}(1989)\citenamefont {Wang},
  \citenamefont {Wu}, \citenamefont {Zhu}, \citenamefont {Shen}, \citenamefont
  {Zhang}, \citenamefont {Yang},\ and\ \citenamefont {Zhao}}]{Wang2}%
  \BibitemOpen
  \bibfield  {author} {\bibinfo {author} {\bibfnamefont {Y.~N.}\ \bibnamefont
  {Wang}}, \bibinfo {author} {\bibfnamefont {J.}~\bibnamefont {Wu}}, \bibinfo
  {author} {\bibfnamefont {J.~S.}\ \bibnamefont {Zhu}}, \bibinfo {author}
  {\bibfnamefont {H.~M.}\ \bibnamefont {Shen}}, \bibinfo {author}
  {\bibfnamefont {J.~Z.}\ \bibnamefont {Zhang}}, \bibinfo {author}
  {\bibfnamefont {Y.~F.}\ \bibnamefont {Yang}},\ and\ \bibinfo {author}
  {\bibfnamefont {Z.~X.}\ \bibnamefont {Zhao}},\ }\bibinfo {title}
  {{U}ltrasonic study on anisotropic elasticity of
  {B}i$_2${S}r$_2${C}a{C}u$_2${O}$_8$ single crystal},\ in\ \href@noop {}
  {\emph {\bibinfo {booktitle} {Beijing International Conference on High
  Temperature Superconductivity}}},\ Vol.~\bibinfo {volume} {22},\ \bibinfo
  {editor} {edited by\ \bibinfo {editor} {\bibfnamefont {Z.~X.}\ \bibnamefont
  {Zhao}}, \bibinfo {editor} {\bibfnamefont {G.~J.}\ \bibnamefont {Cui}},\ and\
  \bibinfo {editor} {\bibfnamefont {R.~S.}\ \bibnamefont {Han}}}\ (\bibinfo
  {publisher} {World Scientific},\ \bibinfo {address} {Beijing},\ \bibinfo
  {year} {1989})\ pp.\ \bibinfo {pages} {426--429}\BibitemShut {NoStop}%
\bibitem [{\citenamefont {{O. Radulescu and T.
  Janssen}}(1997)}]{radulescu1997}%
  \BibitemOpen
  \bibfield  {author} {\bibinfo {author} {\bibnamefont {{O. Radulescu and T.
  Janssen}}},\ }\href@noop {} {\bibfield  {journal} {\bibinfo  {journal} {J.
  Phys. A: Math. Gen.}\ }\textbf {\bibinfo {volume} {30}},\ \bibinfo {pages}
  {4199} (\bibinfo {year} {1997})}\BibitemShut {NoStop}%
\bibitem [{\citenamefont {Finger}\ and\ \citenamefont
  {Rice}(1983)}]{Finger1983}%
  \BibitemOpen
  \bibfield  {author} {\bibinfo {author} {\bibfnamefont {W.}~\bibnamefont
  {Finger}}\ and\ \bibinfo {author} {\bibfnamefont {T.~M.}\ \bibnamefont
  {Rice}},\ }\href {https://doi.org/10.1103/PhysRevB.28.340} {\bibfield
  {journal} {\bibinfo  {journal} {Phys. Rev. B}\ }\textbf {\bibinfo {volume}
  {28}},\ \bibinfo {pages} {340} (\bibinfo {year} {1983})}\BibitemShut
  {NoStop}%
\end{thebibliography}%

\end{document}